\documentclass[
reprint,
nofootinbib,
 amsmath,amssymb,
 aps,
prd,
]{revtex4-1}
\usepackage{lipsum}
\usepackage{dcolumn}
\usepackage{bm}
\usepackage{color}
\usepackage{graphicx,amsmath,amssymb}
\usepackage{tikz}
\usetikzlibrary{matrix}

\newcommand{\eqnsplit}[1]{\begin{align}\begin{split}#1\end{split}\end{align}}
\newcommand{\eqn}[1]{\begin{align}#1\end{align}}

\definecolor{awesome}{rgb}{1.0, 0.13, 0.32}
\definecolor{applegreen}{rgb}{0.55, 0.71, 0.0}
\definecolor{darkpastelgreen}{rgb}{0.01, 0.75, 0.24}
\definecolor{azure(colorwheel)}{rgb}{0.0, 0.5, 1.0}
\definecolor{fluorescentyellow}{rgb}{0.8, 1.0, 0.0}
\definecolor{guppiegreen}{rgb}{0.0, 1.0, 0.5}
\definecolor{inchworm}{rgb}{0.7, 0.93, 0.36}
\definecolor{richelectricblue}{rgb}{0.03, 0.57, 0.82}
\definecolor{springgreen}{rgb}{0.0, 1.0, 0.5}
\definecolor{mediumcandyapplered}{rgb}{0.89, 0.02, 0.17}

\usepackage{hyperref}
\hypersetup{
    bookmarks=true,         
    unicode=false,          
    pdftoolbar=true,        
    pdfmenubar=true,        
    pdffitwindow=false,     
    pdfstartview={FitH},    
    pdftitle={My title},    
    pdfauthor={Author},     
    pdfsubject={Subject},   
    pdfcreator={Creator},   
    pdfproducer={Producer}, 
    pdfkeywords={keyword1, key2, key3}, 
    pdfnewwindow=true,      
    colorlinks=true,       
    linkcolor=richelectricblue,          
    citecolor=darkpastelgreen,        
    filecolor=richelectricblue,      
    urlcolor=darkpastelgreen           
}

\begin{document}
\title{Entanglement entropy on finitely ramified graphs}
\author{Ibrahim Akal}
\affiliation{Theory Group, DESY, Notkestra{\ss}e~85, D-22607 Hamburg, Germany}
\email{ibrahim.akal@desy.de}

\date{\today}
\begin{abstract}
We compute the entanglement entropy in a composite system separated by a finitely ramified boundary with the structure of a self-similar lattice graph. We derive the entropy as a function of the decimation factor which determines the spectral dimension, the latter being generically different from the topological dimension. For large decimations, the graph becomes increasingly dense, yielding a gain in the entanglement entropy which, in the asymptotically smooth limit, approaches a constant value. Conversely, a small decimation factor decreases the entanglement entropy due to a large number of spectral gaps which regulate the amount of information crossing the boundary. In line with earlier studies, we also comment on similarities with certain holographic formulations. Finally, we calculate the higher order corrections in the entanglement entropy which possess a log-periodic oscillatory behavior.
\end{abstract}

\maketitle

\section{Introduction}
Entanglement is one of the most intriguing features of the physical world.
It lays the foundation of quantum information theory as well as other branches of physics such as quantum computing and cryptography \cite{bennett1993teleporting,bennett2000quantum,nielsen2000quantum}.
It has an important role in condensed matter physics, particularly in strongly correlated systems \cite{zanardi2002fermionic,corboz2010simulation}, the latter providing a natural ground for the emergence of collective quantum phenomena.
Entanglement is also responsible for long-range correlations in the ground state of a given quantum system and is therefore fundamental for understanding the nature of quantum phase transitions \cite{osborne2002entanglement,sachdev2007quantum,gagatsos2013mutual}. Recently, the time evolution of entanglement production has been investigated as well, see e.g. \cite{Hackl:2017ndi}.

Entanglement entropy (EE) can be thought of as a measure of quantum entanglement within a system. In the specific case of
an isolated composite system divided into two subregions separated by a codimension-2 surface, $\Sigma$, EE is the measure of the strength of correlations that cross $\Sigma$. 
It was found that this entropy is proportional to the area of $\Sigma$ and
depends on the UV cutoff which regulates the short distance behavior
in the quantum system -- known as the famous area law \cite{eisert2010colloquium}.
The EE for a massless scalar field in a $D$-dimensional spacetime
is of the typical form $S_E \simeq A(\Sigma)\epsilon^{-D+2}$,
where $A(\Sigma)$ is the area of $\Sigma$ and $\epsilon$ denotes the cutoff.

Being a fundamentally nonlocal quantity, EE depends on the entire reduced density matrix of a given subsystem. Therefore, it is extremely challenging to detect EE in an actual experiment.
However, there are recent proposals for the measurement of EE in noninteracting as well as interacting systems
\cite{klich2009quantum,song2012bipartite,abanin2012measuring}.

EE also plays and important role in holography \cite{hooft1993dimensional}, specifically in the AdS/CFT correspondence
\cite{maldacena1999large}
where it serves as a characterization of the conformal boundary.
It has been found that EE on the boundary has
a geometric correspondence in the bulk in terms of the area of a minimal \cite{ryu2006aspects,ryu2006holographic} -- in temporal settings, an extremal \cite{Hubeny:2007xt} -- surface.

More recently, quantum entanglement has been shown to govern the emergence of classical, smoothly connected spacetime \cite{Swingle:2009bg,VanRaamsdonk:2009ar,VanRaamsdonk:2010pw,Rangamani:2016dms}.
Such ideas even led to approach gravity as a consequence of the quantum information theoretic properties of spacetime \cite{Susskind:2014rva,Aaronson:2016vto,Susskind:2018tei}.

In the present paper, we study the entanglement entropy in a composite system with a separating boundary finitely ramified in form of a self-similar lattice graph.
Our analytic results depend on the decimation factor characterizing the underlying graph topology.
Self-similar structures, also referred to as deterministic fractals, appear
in various branches of physics ranging from condensed matter \cite{gefen1980critical,havlin1987diffusion,sornette1998discrete,dal2012deterministic} to quantum field theory \cite{ambjorn2001dynamically,ambjorn2004emergence,lauscher2005fractal,hovrava2009spectral,Calcagni:2009kc,reuter2011fractal}.
More precisely, they have been applied to study phase transitions in critical many body quantum systems by noticing the underlying scale invariance at the critical point \cite{nauenberg1975scaling,hughes1981random,domb2000phase}, which has notable resemblance with the renormalization group approach \cite{suzuki1983phase}.
Beyond that, there is a close similarity between quantum fields on self-similar structures and quantum graphs/networks with applications in various areas including dynamical systems, superconductivity, quantum chaos and mesoscopic physics \cite{kostrykin1999kirchhoff,kottos1997quantum,kuchment2005quantum,gnutzmann2006quantum}. More recently, lattice graphs have also been employed as tensor networks to study critical quantum phenomena \cite{lee2017tensor}.

The present paper is organized as follows:
in section \ref{sec:entropies}, we introduce the basic quantities and sketch the computation strategy. In section \ref{sec:surface}, we describe the setup under consideration. We discuss the graph structure
and provide the associated spectral quantities.
Section \ref{sec:ee} is the main part of this paper. We compute the EE and discuss our results.
The paper is finalized with a brief conclusion in section \ref{sec:conc}.

\section{Entangled subsystems}
\label{sec:entropies}
Consider a pure quantum state
in a spacelike region with degrees of freedom
taken to be located inside certain subregions.
Let $\Sigma$ be the surface separating the original region
in two complementary subregions, $\mathsf{S}_\text A$ and $\mathsf{S}_\text B$.
The resulting global system is the unification $\mathsf{S}_\text A \cup \mathsf{S}_\text B$
and the Hilbert space $\mathcal{H}$ reads $\mathcal{H} = \mathcal{H}_\text A \otimes \mathcal{H}_\text B$.
The density matrix, $\rho$, for the global non-degenerate ground state at zero temperature has zero entropy.
Tracing, for instance, over the degrees of freedom in $\mathsf{S}_\text A$,
the local, reduced density matrix associated with $\mathsf{S}_\text B$ is $\rho_\text B = \mathrm{Tr}_\text A \rho$.
The following statistical quantity
\eqn{
S_\text B = - \mathrm{Tr}\ \rho_\text B\ln\rho_\text B
\label{eq:vNeumann}
}
is called the von Neumann entropy and
\eqn{
S_\text B^{(\alpha)} = \frac{1}{1-\alpha} \ln\ \mathrm{Tr}_\text B\rho_\text B^\alpha
\label{eq:Renyi}
}
is referred to as the $\alpha$-th moment R{\'e}nyi entropy \cite{Calabrese:2009qy}.
Since $\rho$ is assumed to be a pure quantum state,
the relations $S_\text A = S_\text B$
and
$S_\text A^{(\alpha)} = S_\text B^{(\alpha)}$
are fulfilled and only determined by the size of $\Sigma$.
Due to this, EE can be defined as $S_{E} = S_{\text A,\text B}$
where $S_{E} > 0$ corresponds to correlations in the global system
crossing $\Sigma$.

It should be noticed that in the case of mixed quantum and classical correlations, $S_E$ cannot be taken as an appropriate EE measure.
On the contrary, in the situation described above,
calculating the R{\'e}nyi entropy for any $\alpha \geq 1$ yields the EE according to
$S_{E} = \lim_{\alpha \rightarrow 1} S^{(\alpha)}$.
Although the latter relation dictates a clear prescription,
computing the trace in \eqref{eq:Renyi}
can be challenging.
A workable strategy relies on the Euclidean path integral formulation and
the replica trick.

Let $\Sigma$ be separating the hypersurface at zero time.
Then, the trace can be related to the effective action $W_\alpha$
evaluated on a Euclidean manifold. More precisely, the manifold is taken as a flat cone $\mathcal{C}_\alpha$ with angle deficit $2 \pi (1-\alpha)$ where $\Sigma$ sits in the conical singularity \cite{Solodukhin:2011zr}.
Once $W_\alpha$ is computed, by assuming the analytical continuation of $n$ to nonintegers the entropy can be obtained from
\eqn{
S_{E} = \lim_{\alpha \rightarrow 1}[\alpha \partial_\alpha - 1] W_\alpha.
\label{eq:S_E}
}

Performing the standard modifications using Frullani's integral,
we end up with the following effective action
\eqn{
W_\alpha = - \frac{1}{2} \int_{\epsilon^2}^\infty \frac{ds}{s} K_\alpha(s).
\label{eq:W(n)}
}
The integrand $K_\alpha$
is the heat kernel trace defined on the Euclidean space with the aforementioned conical singularity.

What remains to be done is the computation of the latter for which,
fortunately, a shortcut does exist.
A space that is the $\alpha$-fold covering of a smooth manifold, i.e. $\mathcal{C}_\alpha$,
can be seen as a direct product of the stationary point of an Abelian isometry generated by the corresponding Killing vector, here $\Sigma$,
and a two-dimensional cone $\mathcal{C}_{2,\alpha}$ with the same angle deficit as above \cite{Solodukhin:2011gn}.

As a result, we may approximate
$\mathcal{C}_\alpha \approx \mathcal{C}_{2,\alpha} \otimes \Sigma$ which
immediately leads to $K_\alpha \approx K_{2} \times K_\Sigma$.
The computation of $K_{2}$ can be done with the help of the Sommerfeld formula.
From now on, we focus on the massless Klein-Gordon field.
In this case, the trace of the heat kernel on $\mathcal{C}_{2,\alpha}$ reads
\eqn{
K_{2} = \frac{\alpha V}{4 \pi s} + \frac{\alpha C(\alpha)}{2}.
}
The function $C(\alpha) = (1-\alpha^2)/(6\alpha^2)$ stems from the residues associated with the corresponding contour integral in the Sommerfeld formula \cite{Solodukhin:2011gn}.
The first term is proportional to the spacetime volume and contributes to the ground state energy of the system.
The relevant contribution for $S_{E}$ is the second term $\propto C(\alpha)$.

\section{Self-similar ramifications}
\label{sec:surface}
In this section we discuss the separating boundary in detail.
As previously commented, we assume a graph structure that is finitely ramified, which means it can be disconnected by removing a finite number of appropriately chosen vertices. The specific graphs we consider are referred to as diamond graphs \cite{gefen1980critical,hambly2010diffusion}.
These are defined as infinitely iterated, self-similar graphs \cite{bajorin2007vibration}.
For any finite iteration a diamond graph can be identified as a quantum graph \cite{harrison2011zeta,dunne2012heat}.

Here we do not consider these constructions as some effective model for a quantum \textit{spacetime} \cite{Calcagni:2009kc,Calcagni:2011kn}. Instead, our goal is to identify how the modified spectral properties on the
graphs are reflected in the ground state EE. On top of that, the notion of an area law as present for a smooth manifold shall be examined.
Also note that diamond graphs belong to the class of constructions for which certain interesting connections to holographic
duals have recently been observed. More details will be discussed in section~\ref{sec:ee}.

Having said this, in addition to the self-similarity, which basically enables the underlying problem to be solved exactly, diamond graphs have
further remarkable properties.
For instance, the trace of the heat kernel on the graph shows a log-periodic oscillatory behavior in its small time asymptotics. This is generic for many deterministic constructions due to the appearance of a whole tower of
poles in the meromorphic plane which can be identified with an associated complex dimension \cite{lapidus2012fractal}.
Notably, even for a finite number of iterations, the heat kernel trace behaves equivalently \cite{dunne2012heat}.

In the ideal case, i.e. for infinitely iterated constructions, there exists a direct relation between different notions of dimensions which exceed the usual topological dimension of a smooth manifold.
These can most compactly be introduced via the relation $d_w d_s = 2d_h$.
The spectral dimension $d_s$
determines the scaling properties of the system's eigenvalues,
although an alternative definition for $d_s$ is possible as will be seen below.
The Hausdorff dimension $d_h$ refers to the spatial scaling properties.
The third quantity, $d_w$, called the walk dimension, has the physical meaning of a diffusion parameter which
generalizes the usual Einstein relation for Brownian motion \cite{rammal1983random}.
For self-similar constructions, one generally has $d_w > 2$ which leads to a slower spread compared to smooth manifolds.

\begin{figure}[h!]
\centering
  \begin{minipage}[b]{0.2\textwidth}
    \includegraphics[width=.7\textwidth]{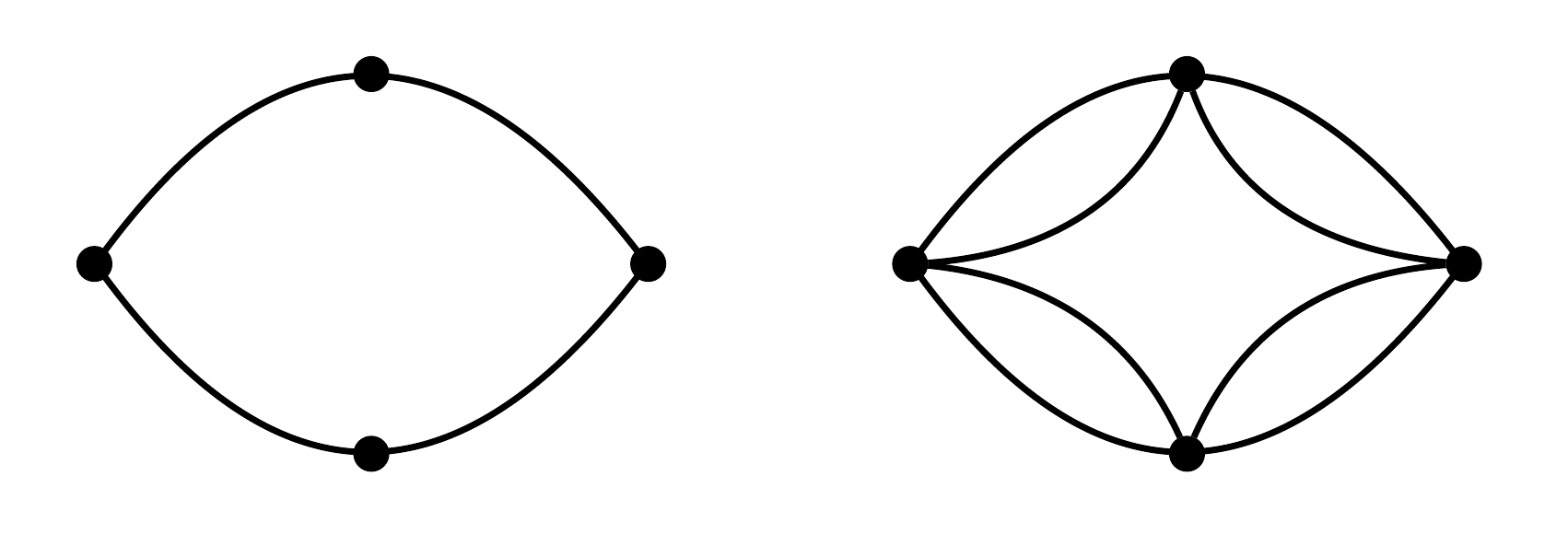}
  \end{minipage}
  \begin{minipage}[b]{0.2\textwidth}
    \includegraphics[width=.76\textwidth]{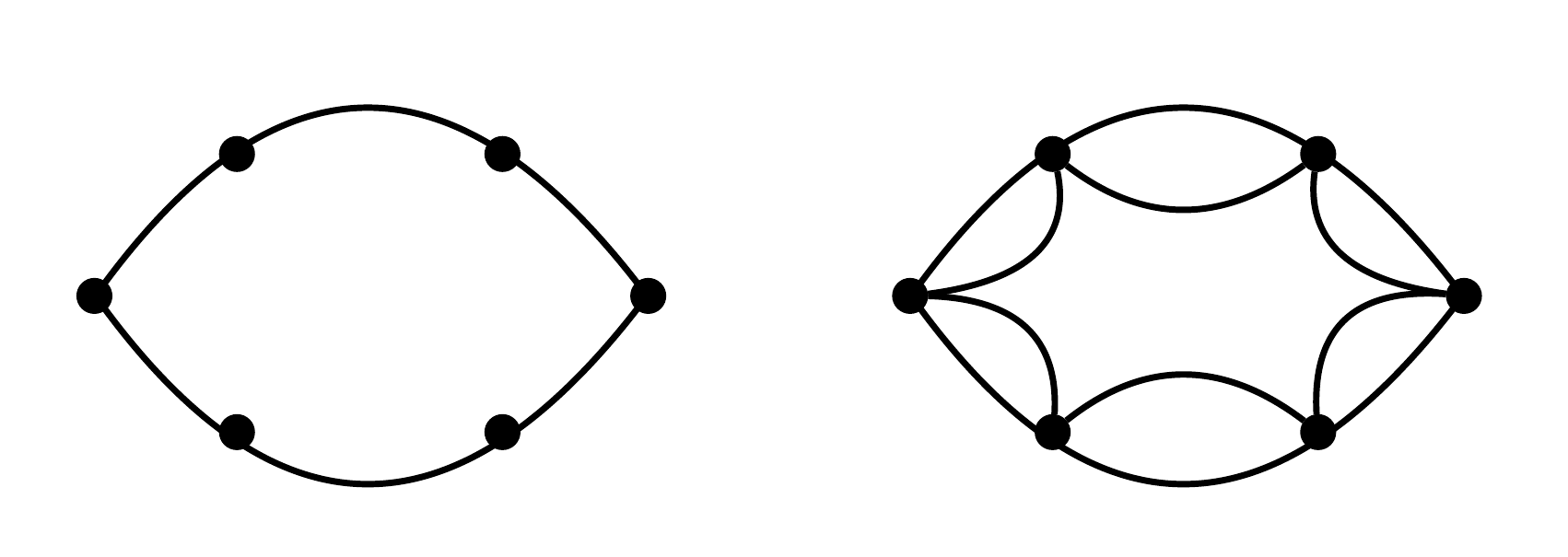}
  \end{minipage}
\caption{Diamond graphs with different topologies $\mathbf{D}_{4,2}$ (left) and $\mathbf{D}_{6,3}$ (right). For both cases, the initial link is shown on the left side with vertices of level $k$. Next to that, the corresponding iterated graphs are depicted including vertices of level $k+1$. The associated Hausdorff dimension $d_h$ and decimation factor $l$ are consistent with $\mathbf{D}_{l^{d_h},l}$.}
\label{fig:diamonds}
\end{figure}

Let us make the current setup explicit: we start with an initial object, called a link. This link has a one-fold branching which splits into two legs where each of the legs is divided into two edges. By doing so, the original link is divided into four edges which are located between the vertices depicted in the left panel in figure~\ref{fig:diamonds}.
With the next iteration step, each of the edges becomes a link itself.
In this way, the number of links for which the initial one is divided into is four whereas,
the decimation factor is two.

The class of diamond graphs can be described by two integer numbers.
We introduce the following notation $\mathbf{D}_{l^{d_h},l}$
for identifying the specific graph topology.
Here we restrict ourselves to graphs for which the number of links the initial one is divided into, $l^{d_h}$, is always twice that of the decimation factor, resulting in the corresponding branching factor $l^{d_h-1}-1=1$.
Thus, the previously described construction with $l=2$ belongs to the class $\mathbf{D}_{4,2}$ with $d_h = 2$, since $d_h = \ln(2l)/\ln(l)$.
The case with $l=3$ is sketched in the right panel in figure~\ref{fig:diamonds}.
All diamond graphs have a fixed walk dimension $d_w = 2$.
Hence, the diffusion is anomaly free or,
in other words, the spectral and Hausdorff dimensions coincide, $d_s = d_h$.
This coincidence is a unique feature that does not affect the characteristic spectral properties.

In what follows, we assume $l \in \mathbb{N}_{> 2}$ corresponding to $1 < d_s < 2$
where $D=(1+d_s) + 1$.
Separating the zero mode, the zeta function can be solved exactly \cite{Akkermans:2009sb},
\eqn{
\zeta(s) = \frac{2 \zeta_R(2s)}{\pi^{2s}} \left[ \frac{1-l^{1 - 2 s}}{1-l^{d_s - 2 s}} \right]
\label{eq:exact-zeta}
}
with $\zeta_R$ denoting the standard Riemann zeta function.
The function $\zeta$ has meromorphic continuation to the complex plane with a whole tower of poles $s_n$
\eqn{
s_n = \frac{d_s}{2} + \frac{i \pi n}{\ln(l)},\quad n \in \mathbb{Z}
\label{eq:poles}
}
such that $d_s$ can alternatively be defined
as the (largest, if other towers exist) real part of the poles.
Using the residues, we can compute the inverse Mellin transform which leads to the following heat kernel trace $K(s)$
\eqn{
K(s) \simeq \zeta_0 + \frac{A_s}{s^{d_s/2}} +  \sum_{n \geq 1} \ldots ,
\label{eq:kernel-trace}
}
where
\eqnsplit{
\zeta_0 &:= \zeta_R(0) \left[ \frac{2 - l^{d_s}}{1 - l^{d_s}} \right],\\
A_s &:=  \frac{1}{2 \ln(l)}  \left[ \frac{\zeta_R(d_s)\Gamma(d_s/2)}{\pi^{d_s}} \right].
\label{eq:defs}
}
Here, $\Gamma$ represents the standard Gamma function. The first expression in \eqref{eq:defs} can be dropped, since
the relevant terms are the $s$ dependent one in \eqref{eq:kernel-trace}.
The second line, $A_s$,
is taken as the spectral area.

Generally, the notion of an area for some graph is not so obvious and one has to
specify an appropriate definition.
It has been shown that random graph states generally satisfy an area law on average \cite{collins2013area}.
For the present case, a possible way out can be achieved by probing the geometry of the region
in terms of a given test \textit{particle} at thermal equilibrium \cite{dowker1978finite,dowker1984finite}.

Starting from the thermodynamic equation, it has been shown that the area on smooth manifolds, $A$, scales as
\eqn{
A \sim \mathrm{Res}_{d/2}[\zeta_s(s)] \Gamma\left( d/2 \right)
}
where $\zeta_s$ denotes the zeta function on the manifold under consideration, see e.g. \cite{akkermans2010thermodynamics,dunne2012heat}.
Here we have used the (rescaled) length $L=1$.
In close analogy, the area on the graph 
can be defined upon replacing $d$ by $d_s$ and $\zeta_s$ by $\zeta$, respectively,
which corresponds (up to some constants) to the definition $A_s$ in \eqref{eq:defs}. Of course, as soon as the graph becomes denser, i.e. when it approaches a smoothly connected space, one recovers the usual area, $A_s \rightarrow A$.

Referring to \eqref{eq:kernel-trace}, we already see that the spectral dimension determines the nonoscillatory part
in the small time asymptotics. The oscillatory terms, represented by the dots, depend on the
walk dimension $d_w = 2$.
For the spectral dimension we find $\lim_{l \rightarrow \infty} d_s = 1$,
corresponding to the rescaled line segment with unit length,
which can be also deduced from the zeta function $\zeta(s)$.
Namely, for $l \rightarrow \infty$ the expression in the brackets 
yields a factor $1/2$ such that $\zeta(s) \rightarrow \zeta_R(2s)/\pi^{2 s}$, which is equivalent to 
the zeta function for the line segment.
Since the diffusion is anomaly free ($d_w = 2$), this leads to the expected smooth limit with $D=2 + 1$.

In the next section we examine how the previous observations get reflected in the EE for the case $\Sigma = \mathbf{D}_{2l,l}$.

\section{Entanglement entropy}
\label{sec:ee}
\subsection{Leading order}
We begin with the non-oscillatory part of the heat kernel trace, dropping
all higher order terms with $n \geq 1$ in \eqref{eq:kernel-trace}.
Using equation~\eqref{eq:W(n)}, the effective action is given by
\eqn{
W_\alpha = \frac{A_s \left(\alpha^2-1\right)}{12 \alpha d_s \epsilon^{d_s}}.
\label{eq:Walpha}
}
Afterwards, applying the formula \eqref{eq:S_E}, we find the main result in this paper,
\eqn{
S_E \simeq \frac{A_s}{d_s \epsilon^{d_s}}
\sim \frac{1}{\epsilon^{d_s}}  \left[ \frac{\zeta_R\left( \frac{\ln(2 l)}{\ln(l)} \right) \Gamma\left( \frac{\ln(2l)}{2\ln(l)} \right)}{2 \ln(2 l)} \right],
\label{eq:S_E-res}
}
the leading term in the EE.

Note that the first expression on the right-hand side of \eqref{eq:S_E-res} directly follows from a sub-Gaussian
estimate for the heat kernel trace \cite{Astaneh:2015gmg}. As an interesting side note, 
the mentioned expression in \eqref{eq:S_E-res}
could provide evidence for a possible dual description of a fractal boundary 
within holographic hyperscaling violating theories. Namely, referring to the Ryu-Takayanagi formula, the leading order contribution
to $S_E$ in the case of such bulk theories has been shown to obey
$S_E \sim (d_\theta \epsilon^{d_\theta})^{-1}$ where $d_\theta$ equals the dimension of the corresponding boundary theory.
Apparently, such a scaling behavior perfectly coincides with the one in \eqref{eq:Walpha} if we allow $d_\theta \leftrightarrow d_s$. An analogy is further supported via a comparison with the associated thermal entropy \cite{Astaneh:2015gmg}.
In addition to these insights, although in a different context, note the recent findings in \cite{Guralnik:2018kwo} which also indicate certain
connections between a fractal geometry and the AdS/CFT correspondence.
These are interesting coincidences which deserve further investigations. 

Apart from the general scaling behavior, in the present case we have derived an explicit expression for $S_E$ depending on the decimation factor which can be seen as a measure for the size of the spectral gaps.
\begin{figure}[h!]
\centering
\includegraphics[width=.45\textwidth]{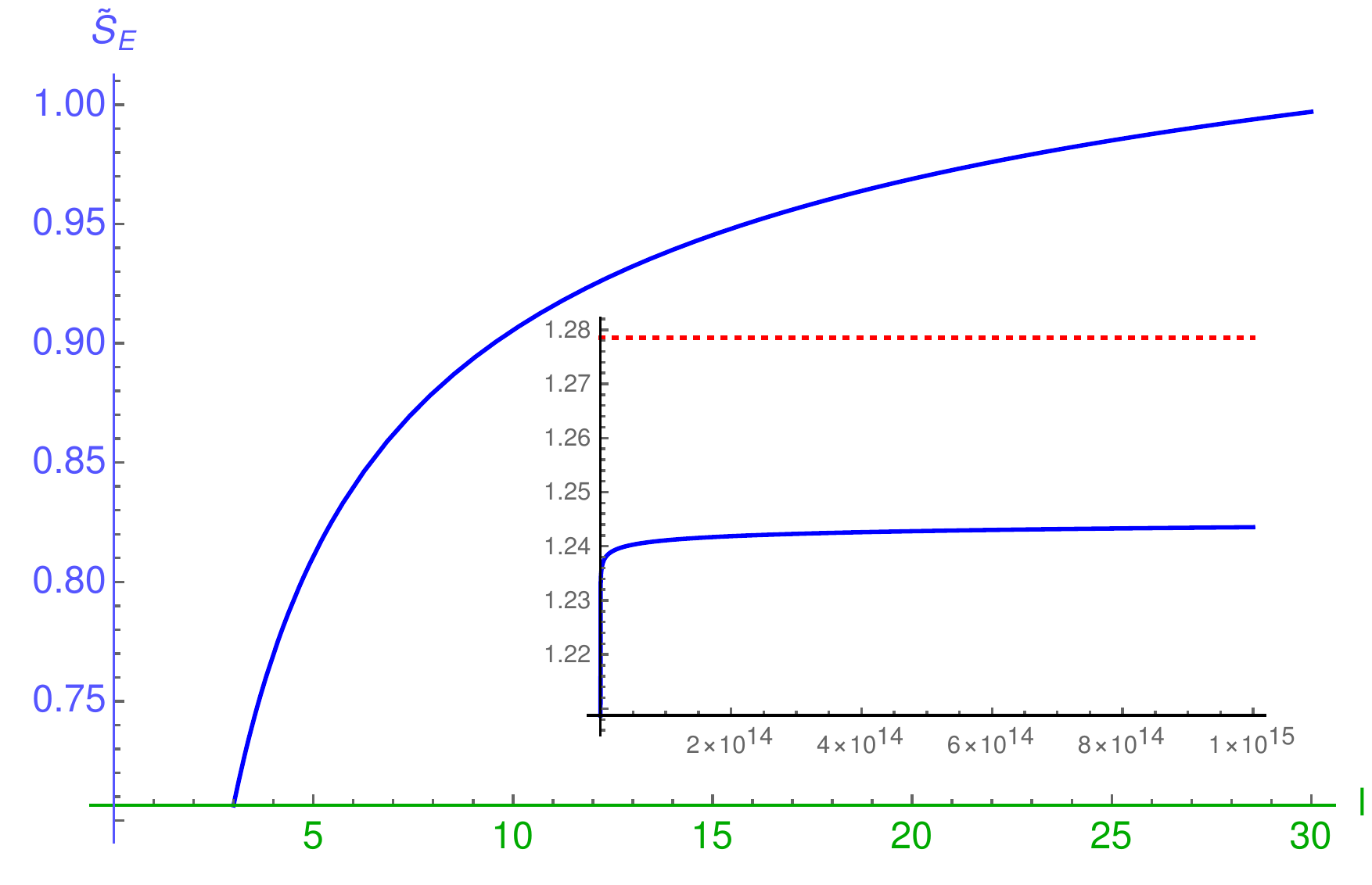}
\caption{$\tilde S_E$ is plotted versus the decimation factor $l$.
The behavior for large $l$ is depicted in the inset. The horizontal red dashed line
at $\tilde S_E \approx 1.28$ is obtained for $l \rightarrow \infty$.}
\label{fig:As-l}
\end{figure}
Note that for $d_s \rightarrow D - 2$ ($\text{codim}(\Sigma) \rightarrow 2$) we regain the usual area law for a smooth manifold.
In figure~\ref{fig:As-l}, $\tilde S_E := S_E \epsilon^{d_s}$ is plotted versus $l$ which
clearly shows that increasing the decimation factor raises the EE.
With sufficiently large values the curve gets flattened and settles at $\tilde S_E \approx 1.28$ marked by the red dashed horizontal line in the inset.
Recall that $l \rightarrow \infty$ is the asymptotically smooth limit where the EE has to be independent of any discrete quantity.
A large decimation factor fills in more gaps
which, subsequently, leads to a larger spread of the entangling degrees of freedom across the separating boundary.
This continues until the boundary becomes perfectly smooth. So, for $l \rightarrow \infty$ the EE only depends on the topological area. which explains the described saturation in figure \ref{fig:As-l}.
On the other hand, for moderate $l$, the effect of the ramifications
manifests itself in a smaller EE.
This observation, for instance, is interesting from the following perspectives.

A small decimation factor yields a separating boundary having large spectral gaps, though it is infinitely iterated.
This implies that quantum information cannot uniformly spread
as on a smooth manifold -- see also \cite{hamma2010entanglement,genzor2016phase} for similar reduction effects.
It is worth stressing that diffusion in disordered systems (e.g. the motion of a particle in a fluid) or percolation phenomena (e.g. the soaking process in a porous medium) can show a variety of interesting properties such as anomalous spread (percolation being distinguished by the possibility of trapping and blocking) and
non-Gaussian probability density
\cite{shante1971introduction}.
Notably, in quantum percolation \cite{mookerjee1995quantum} an additional complexity appears due to Anderson localization \cite{anderson1958absence}, as a consequence of quantum interference at the nodes. This
effect has recently been observed on random fractal lattices resulting in a strong dependence on the spectral dimension \cite{kosior2017localization}.
Referring to our findings on the diamond graph, one may therefore consider analogies to quantum percolation.
Recall that the spread in our case is anomaly free by construction.
Hence, the described reduction of the EE unveils itself as a consequence of the spectral gaps.
Moreover, the observed behavior reflects the interesting relation between the level of smoothness and
information flow. This may in particular be relevant for mesoscopic systems with some disordered structure \cite{havlin1987diffusion,kuchment2002graph,comtet2005functionals}.
Note that we may interpret $l$ as an effective measure for the number of vertices
on the graph \cite{akkermans2010thermodynamics}. 

In summary, the curve for $\tilde S_E$ reveals a clear trend with respect to the decimation factor.
It would be particularly interesting to identify how $S_E$ declines with increasing the size of the spectral gaps.
In the present setup, the smallest decimation factor we can apply is limited to $l=3$.
For $l < 3$, there is no possibility to find an exact expression for $\zeta(s)$.
However, the present example provides substantial evidence that lowering $S_E$
should be possible in an appropriate setting.

Finally, note that 
even if the spectral volume $A_s$ can be reduced to sufficiently small values,
the entropy $S_E$ still remains divergent in the deeply UV regime.
Such a behavior is naturally predicted by the microscopic structure of any local quantum field theory
having a large number of degrees of freedom at high energies, see e.g. \cite{reeh1961bemerkungen,srednicki1993entropy}.
An additional dimensional flow can render the entropy finite \cite{Amelino-Camelia:2017pdr} -- see also
\cite{Pagani:2018mke}
for recent findings.
Such insights could be of interest for multi-self-similar graphs. In fact, it has been observed that the entropy can extract universal information about a system's inherent multifractal structure \cite{chen2012renyi}.

\subsection{Higher orders}
\begin{figure}[h!]
\centering
\includegraphics[width=.23\textwidth]{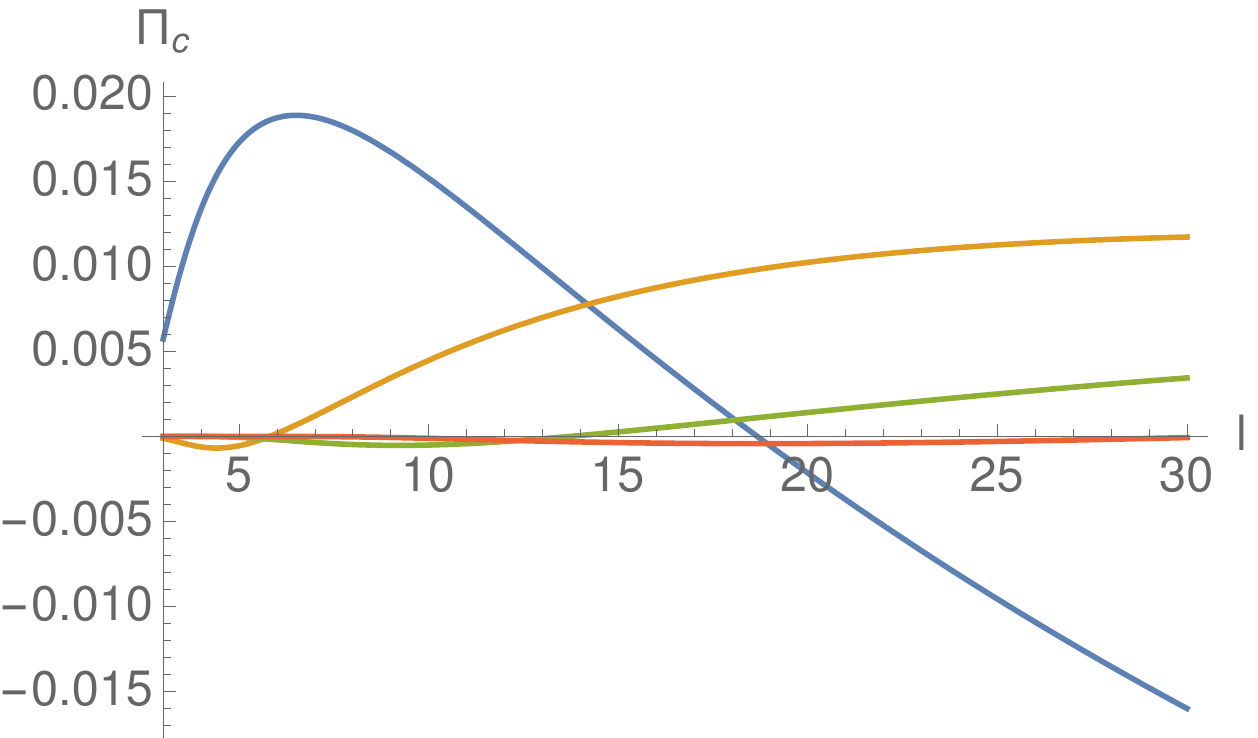}
\includegraphics[width=.245\textwidth]{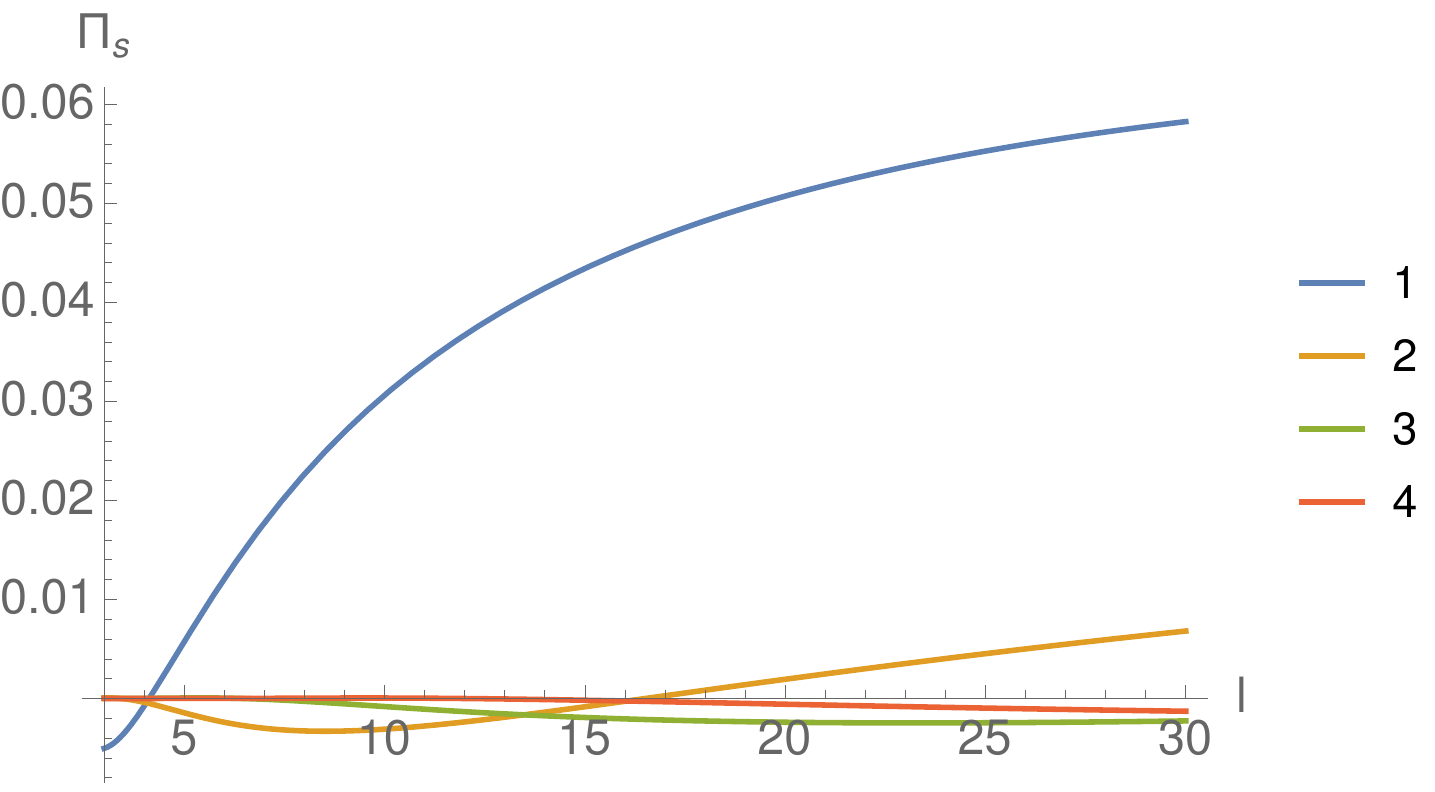}
\caption{$\Pi_c$ (left) and $\Pi_s$ (right) are plotted versus $l$.
The numbers in the legends correspond to $n$ introduced in \eqref{eq:poles}.}
\label{fig:Pi_c,s}
\end{figure}
As discussed in section~\ref{sec:surface}, the heat kernel trace receives log-periodic corrections
from the imaginary part of the poles $s_n$ for any finite decimation factor.
Log-periodic oscillations have been observed in several other contexts as a consequence of
discrete scale invariance associated with critical phenomena \cite{gluzman2002log}.
In the present case, the exact trace result is \cite{dunne2012heat}
\begin{align}
K(s) \simeq
\zeta_0
&+ \frac{A_s}{s^{d_s/2}}
\bigg[ 1
+ \sum_{n \geq 1}
\Delta_{\Re,n} \cos\left(\ln\left(s^{n \pi/\ln(l)}\right)\right)\nonumber\\
&+  \sum_{n \geq 1}
\Delta_{\Im,n} \sin\left(\ln\left(s^{n \pi/\ln(l)}\right)\right)
\bigg].
\label{eq:K(s)osc}
\end{align}
Using the latter, we find the following EE when
including the higher order terms
\eqnsplit{
S_E \simeq \frac{A_s}{d_s \epsilon^{d_s}}
\bigg[1
&+ \sum_{n \geq 1} \Pi_c \cos\left( \frac{\pi \ln(\epsilon^2)}{\ln(l)} \right)\\
&+ \sum_{n \geq 1} \Pi_s \sin\left( \frac{\pi \ln(\epsilon^2)}{\ln(l)} \right)
\bigg].
\label{eq:S_En}
}
The prefactors $\Pi_c,\Pi_s$ are of the form
\eqn{
\Pi_c := \frac{\Delta_{\Re,n} + \frac{2 \pi \Delta_{\Im,n}}{d_s \ln(l)}}{1 + \left(\frac{2 \pi}{d_s \ln(l)}\right)^2},\quad
\Pi_s := \frac{\Delta_{\Im,n} - \frac{2 \pi \Delta_{\Re,n}}{d_s \ln(l)}}{1 + \left(\frac{2 \pi}{d_s \ln(l)}\right)^2}
\label{eq:Pi_c,s}
}
where we have defined
\eqnsplit{
\Delta_{\Re,n} &:= \frac{2 \pi^{2 s_0}}{\zeta_R(2 s_0)\Gamma(s_0)}\Re\left(\frac{\zeta_R(2 s_n)\Gamma(s_n)}{\pi^{2 s_n}}\right),\\
\Delta_{\Im,n} &:= \frac{2 \pi^{2 s_0}}{\zeta_R(2 s_0)\Gamma(s_0)}\Im\left(\frac{\zeta_R(2 s_n)\Gamma(s_n)}{\pi^{2 s_n}}\right).
\label{eq:Deltas}
}
Starting with the zeta function $\zeta(s)$ in \eqref{eq:exact-zeta}, the trace in the limit $l \rightarrow \infty$ simplifies to
\eqn{
K(s) \simeq
\zeta_R(0)
+ \frac{A}{s^{d/2}}.
}
Observe that the correction terms in \eqref{eq:S_En} turn out to be UV finite, cf. \cite{Astaneh:2015gmg}, which
is expected to hold for other deterministic self-similar constructions as well.
For illustrative reasons, we have plotted
the prefactors $\Pi_c,\Pi_s$ in
figure~\ref{fig:Pi_c,s}, for
integers $n \in \{1,2,3,4\}$.
A substantial variation between
the different orders is present where the peak strength decreases with $n$.
For any fixed cutoff the oscillations become increasingly weakened for sufficiently large $l$.

\section{Conclusion}
\label{sec:conc}
We have computed the entanglement entropy (EE) across a finitely ramified boundary with the structure of a self-similar lattice graph.
We have found that increasing the topological decimation factor leads to an increase in the EE. In the limit where the decimation factor approaches infinity, the EE settles at a constant value. This is the case in which the separating boundary becomes a perfectly smooth line and hence independent of any discrete quantity. Accordingly, the entangling degrees of freedom are maximally spread
and the EE satisfies the known area law on a smooth manifold.

For small decimation factors, the EE drops substantially. In this case, the size of the spectral gaps increases. Large gaps hinder the spread of the degrees of freedom and, thus, the information flow across the boundary. Resorting to an appropriate definition of the area relying on the thermodynamic properties of a boson on the graph, we have seen that in the presence of these spectral gaps the EE still obeys an area law.

In addition,
we have shown that higher order corrections, as a remnant of the underlying graph symmetry, result in characteristic log-periodic oscillations.

A potential extension of our results would be the computation of the mass induced logarithmic terms in the EE
to ascertain the related \textit{central charges} of the underlying graphs. Such an endeavor would be exciting due to the presence of a similar scaling behavior found in certain holographic formulations.

\section{Acknowledgements}
I would like to thank Zhanybek Alpichshev, Benjamin Bahr and Jim Talbert for valuable discussions and helpful comments on the manuscript.
I acknowledge the support of the SFB 676.

\bibliography{article_bib}
\end{document}